\documentstyle[pre,aps,preprint,amstex,epsfig]{revtex} 
\begin{document}
\title{Phase diagram of an Ising model with long-range frustrating 
interactions: a theoretical analysis.}

\author{M. Grousson$^1$, G. Tarjus$^1$ and P. Viot$^1,^2$}\address{$^1$
Laboratoire de  Physique Th{\'e}orique des  Liquides\\  Universit{\'e}
Pierre  et Marie  Curie   4,   place Jussieu  75252 Paris  Cedex  05
France}\address{$^2$
Laboratoire de
Physique  Th{\'e}orique,\\ Bat.  210,   Universit{\'e} Paris-Sud 91405  ORSAY
Cedex France} 
\maketitle

\begin{abstract}
We  present a theoretical study of  the phase diagram  of a frustrated
Ising  model  with nearest-neighbor   ferromagnetic  interactions  and
long-range   (Coulombic)  antiferromagnetic interactions.  For nonzero
frustration,  long-range ferromagnetic   order is  forbidden,  and the
ground-state of  the  system  consists  of  phases  characterized   by
periodically modulated  structures.  At finite temperatures, the phase
diagram is calculated within  the mean-field approximation.  Below the
transition  line that separates the  disordered and the ordered phases,
the frustration-temperature phase diagram  displays an infinite number
of ``flowers'',  each  flower being  made   by an infinite   number of
modulated phases generated   by  structure combination     branching
processes.  The specificities introduced  by the long-range nature  of
the  frustrating interaction  and the   limitation  of the  mean-field
approach are finally discussed.
\end{abstract}

\section{Introduction}\label{sec:introduction}

There  are many physical examples  in which a short-ranged tendency to
order is opposed by  a long-range frustrating interactions: In diblock
copolymers formed by two mutually incompatible polymer chains attached
to each other, the repulsive short-range forces  between the two types
of components tend  to induce phase  separation of the melt, but total
segregation is  forbidden  by the covalent bonds  that link  the subchains
together. A microphase  separation transition  occurs instead at   low
enough   temperature,   and the  system  then   forms   phases with  a
periodically modulation  of  structures rich in  one  component or the
other,  such as lamellar, hexagonal,  or cubic phases\cite{FH87}. In a
similar way,  self-assembly in  water-oil-surfactant mixtures results
from the competition between the short-range tendency of water and oil
to  phase separate and the stoichiometric  constraints generated by the
presence   of surfactant   molecules,   constraints that  act   as the
electroneutrality  condition  in    a system   of  charged  particles
\cite{WCS92,DC94,WCC95}. The same kind of  physics also arises in quite
different  fields.    For   instance,   stripe   formation in    doped
antiferromagnets   like cuprates  has been  ascribed   to a frustrated
electronic  phase separation, by which  a strong local tendency of the
holes to  phase separate  into a  hole-rich ``metallic'' phase  and  a
hole-poor  antiferromagnetic phase is   prohibited by  the  long-range
Coulombic repulsion between   the holes   \cite{GD82,CKNE98}.  A  last
example is provided  by the structural  or  topological frustration in
glass-forming  liquids: the  dramatic slowing  down  of the relaxation
that  leads to the glass formation   has been interpreted as resulting
from the presence of frustration-limited domains whose formation comes
from  the  inability  of  the   locally  preferred arrangement of  the
molecules in  the   liquid  to tile   space  periodically\cite{KTK96};
topological   frustration   may   also    lead   to    low-temperature
defect-ordered phases such as the Frank-Kasper
\cite{C97} phases in bimetallic systems.

A    coarse-grained    description    of      the  above     mentioned
situations\footnote{additional examples include  cross-linked  polymer
mixture, interpenetrating networks, ultra-thin films} involves lattice
or  continuum  models   with  competing   short-range and    Coulombic
interactions.  The purpose of  the present work  is to study the phase
diagram   of  such  a model, namely    the  Coulomb  frustrated  Ising
ferromagnet  in  which  Ising  spins placed  on  a 3-dimensional cubic
lattice interact via both nearest-neighbor ferromagnetic couplings and
long-range Coulomb-like  antiferromagnetic    terms.   the model    is
introduced in more  detail  in section ~\ref{sec:coulombfrus} and  its
ground states as a function of the frustration  parameter, i.e. of the
ratio of     the  antiferromagnetic coupling     strength  over    the
ferromagnetic one,  is studied in section ~\ref{sec:phase-diagram-at}.
In  section    ~\ref{sec:mean-field-theory},     we investigate    the
finite-temperature  phase  diagram in   the mean-field  approximation.
Finally, the  effect   of the  long-range  nature  of the  frustrating
forces, when comparing the phase behavior of  the prototypical model
with competing,  but short-ranged interactions, the axial next-nearest
neighbor  Ising (ANNNI)  model\cite{E61,SF79,S88},   as  well   as  the
limitations   of the  mean-field  approach   are discussed  in section
~\ref{sec:discussion}.

\section{The Coulomb frustrated Ising ferromagnet}\label{sec:coulombfrus}
The model is described by the Hamiltonian
\begin{equation}\label{eq:1}
H= - J \sum_{<ij>}S_iS_j +\frac{Q}{2}\sum_{i\neq j}v({\bf r}_{ij})S_iS_j,
\end{equation}
where,   $J,Q>0$  are the  ferromagnetic   and antiferromagnetic  coupling
strengths, $S_i=\pm1$ are Ising spin variables placed  on the sites of a
three-dimensional  cubic lattice, $<ij>$   denotes a sum restricted to
nearest neighbors, ${\bf r}_{ij}$ is the vector joining sites i and j,
and $v({\bf r})$ represents a Coulomb-like   interaction term with   $v({\bf
r})_\sim\frac{1}{|{\bf  r}|}$when  $|{\bf r}|\to\infty$.(In all the paper,  the
lattice spacing is taken as the unit length).

In  addition  to  considering   the   true  Coulombic  term,   $v({\bf
r})=\frac{1}{|{\bf  r}|}$,   we have   also  studied, for mathematical
convenience in the analytical treatment, an expression of $v({\bf r})$
in terms  of the lattice Green's function   that satisfies the Poisson
equation on the three-dimensional  cubic  lattice. In the latter  case,
${v\bf (r)}$  is then  simply given,  up  to  a multiplying  factor of
$4\pi$,   as  the inverse   Fourier   transform of the  inverse lattice
Laplacian,
\begin{equation}\label{eq:2}
v({\bf r})=\frac{4\pi}{N}\sum_{{\bf
k}}\frac{\exp(-i{\bf k}{\bf r})}{2\sum_{\alpha=x,y,z}(1-\cos(k_{\alpha}))},
\end{equation}
where  $N$  is the   number  of lattice  sites and   the sum over  ${\bf
k}=(k_x,k_y,k_z)$ is restricted to the first Brillouin zone. For large
$|{\bf r}|$     the     lattice   Green's   function     behaves    as
$1/(4\pi|{\bf r}|)$       so  that      the     expression     in
Eq.~(\ref{eq:2}) has the  proper asymptotic behavior. In
practice, even at  the next-neighbor distance, the  difference between
the true Coulombic form   and Eq.~(\ref{eq:2}) is  very
small \cite{K89}. One  has however to  be careful about two points. The
first       one     is    that     the   $v({\bf r})$     defined   in
Eq.~(\ref{eq:2})  has  a   nonzero, finite    value  at
${\bf r}=0$, $v({\bf 0})=0.25273100986\ldots$, a  value  that must    of
course    be       excluded when  considering     the    Hamiltonian  in
Eq.~(\ref{eq:1}).\footnote{The     correction   term    involving
$v({\bf 0})$ in Eq.~(\ref{eq:4})  has been omitted  in previous papers
\cite{VT98,NRKC99}, but it leads  only to very small corrections  when
the parameter $Q/J$   is small.}  For instance,  Eq.~(\ref{eq:1})
when expressed in Fourier space, can be written as
\begin{equation}\label{eq:3}
H=\frac{J}{2}\sum_{{\bf
k}}\hat{V}({\bf{k}})|\hat{S}({\bf{k}})|^2
\end{equation}
where 
\begin{equation}\label{eq:4}
\hat{V}({\bf k})=-2\sum_{\alpha =x,y,z}\cos(k_\alpha)+\frac{4\pi
Q}{J}\left[\frac{1}{2\sum_{\alpha=x,y,z}(1-\cos(k_\alpha))}-v({\bf 0})\right]
\end{equation}
and $\hat{S}({\bf{k}})$ is the lattice Fourier transform of the Ising
spin variable $S_i$.

To  assess the quantitative difference between  the Coulombic form and
that involving the  lattice Green's function,  we have  calculated the
energy of the system in  several periodic configurations of the spins.
For periodic  lamellar patterns of large width  $m$, the difference is
negligible,    but  it  increases when  $m$    decreases.  For a  N{\'e}el
antiferromagnetic state,  the   Coulombic energy  is related   to  the
Madelung  constant and is equal to   $-0.873782\ldots$, whereas the lattice
Green's function  expression gives  $-1.064356992$.  The  second point
worth  mentioning is that the $v({\bf  r})$  expressed in terms of the
lattice  Green's  function has the   same   discrete symmetry as   the
nearest-neighbor  ferromagnetic interaction whereas the true Coulombic
form has continuous rotational  symmetry instead. The consequences are
negligible for  the   ground-state and mean-field   analyses,  but may
become important for the  finite-temperature behavior of the system in
other approximations\cite{CEKNT96}.

\section{Phase diagram at zero temperature}\label{sec:phase-diagram-at}

In the absence of the  antiferromagnetic interaction ($Q=0$) the model
reduces to the standard Ising ferromagnet, and the ground state of the
system is   obtained when all  spins  are  aligned in  a ferromagnetic
state.  Oppositely, when the  ferromagnetic interaction is set to zero
($J=0$), the  model  is equivalent to a  Coulomb  lattice gas  and the
ground-state  of the  system is a   N{\'e}el antiferromagnetic state. When
$Q\neq  0$,   the Coulombic  interaction  prevents  the existence   of a
ferromagnetic phase,  and   in  the thermodynamic  limit,  the   total
magnetization (charge) is constrained to be zero. Instead, phases with
modulated order, i.e., with  periodic patterns of ``up''  and ``down''
spins subject to the constraint of  zero magnetization, are formed. We
have  studied  these phases  both   analytically with  the  long-range
interactions modeled by the  lattice Green's function  and numerically
with the true Coulombic form.

For small values of the  frustration parameter $Q/J$, the ground state
consists   of  lamellar phases    in  which  parallel  planes  of
ferromagnetically  aligned spins form a   periodic structure along the
orthogonal direction.   The  system in such state  can   mapped onto a
finite one-dimensional system  of  length $2m$ where  $m$  denotes the
width of the lamellae\cite{MWRD95,LEFK94}.

The short-range ferromagnetic contribution to the energy per spin of a
lamellar phase can be  readily calculated, and one gets
\begin{equation}\label{eq:5}
E_{SR}=-J\left(3-\frac{2}{m}\right).
\end{equation}

The Coulombic  energy due to  the  long-range competing forces can  be
calculated in reciprocal space by using the expression in terms of the
inverse lattice Laplacian (see section ~\ref{sec:coulombfrus}).  For a
lamellar phase  of period $2m$ the  wave-vectors to be considered have
only one nonzero component  that takes the values $k=\pi(2n+1)/m$  with
$0\leq n <  m-1$. Correspondingly, the  lattice Fourier transform of the
Ising                           spin                         variable,
$\hat{S}({\bf k})=\frac{1}{\sqrt{N}}\sum_{i=1}^{N}S_i\exp(i{\bf  k}{\bf  r}_i)
$, where N is the total number of  lattice site, has its modulus given
by
\begin{equation}\label{eq:6}
|\hat{S}({\bf k})|=|\hat{S}( k)|=\frac{\sqrt{N}}{m\sin(k/2)}.
\end{equation}
Using then the identities
\begin{eqnarray}
\sum_{n=0}^{m-1}\frac{1}{\sin(\pi(2n+1/(2m))^2}&=&m^2,\\\label{eq:7}
\sum_{n=0}^{m-1}\frac{1}{\sin((\pi(2n+1/(2m))^4}&=&
\frac{m^4}{3}+\frac{2m^2}{3},\label{eq:8}
\end{eqnarray}
one finds  that  the sum  rule $  \sum_{k}|\hat{S}(k)|^2=N$ is properly
satisfied and that the Coulombic energy per site, $E_c$, is equal to 
\begin{eqnarray}
E_c&=&Q \left[\sum_{n=0}^{m-1} \frac{\pi  }{2\sin(\pi(2n+1)/(2m))^4}-2\pi v(0)\right]\\
&=& Q\left[\frac{\pi m^2}{6}+\pi\left(\frac{1 }{3}-2v(0)\right)\right]
\label{eq:9}.
\end{eqnarray}

In   appendix~\ref{sec:calc-coul-energy},  we   show  that  the   same
expression  for the Coulombic energy  is obtained  when performing the
calculation in real space with an effective one-dimensional potential,
this  latter   includes a  convergence   factor that   helps  handling
conditionally convergent sums appearing  as a result of the long-range
nature of the forces  and  that is taken  to  zero at  the end of  the
calculation.   It  is worth mentioning   that the above calculation in
reciprocal  space implicitly assumes   that the contribution from  the
  $k=0$ term is zero;  physically,  this means that the  periodic
system is  embedded by  a medium  with infinite  dielectric  constant.
(For a more   detailed  discussion, see Ref\cite{FS96}.)  Since    for
simulations  on ionic    systems,   a similar  choice    is  generally
adopted\cite{FS96}, both numerical and analytical calculations for our
model have been performed using such metallic boundary conditions.

\begin{figure}
\begin{center}

\resizebox{8cm}{!}{\includegraphics{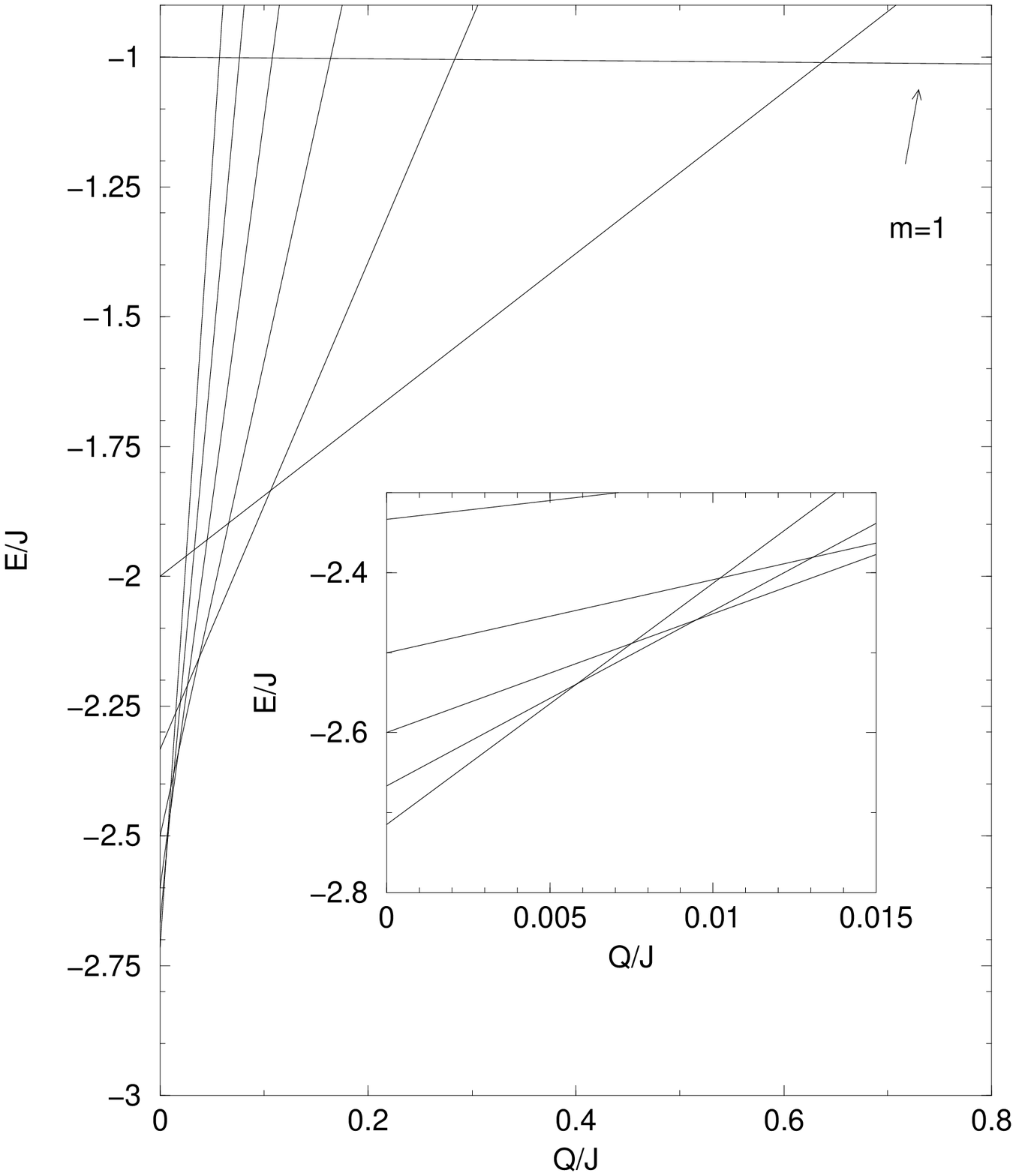}}
\caption{Ground-state energy of the lamellar phases with $m=1,2,..7$
as  a function of  the frustration  parameter $Q/J$ for the inverse
lattice Laplacian expression of $v({\bf r})$.  When
$Q/J$ goes  to zero,  the period  of the  lamellar phases increases.  The
inset zooms in on the region of vanishing frustration parameters which
corresponds to the left down region of the figure.}\label{fig:1}
\end{center}\end{figure}

Figure.~\ref{fig:1} shows  the total energy per  spin, $E/J$, which is
the sum of the two contributions in Eq.~(\ref{eq:5}) and ~(\ref{eq:9})
for the inverse  lattice Laplacian expression, as  a function of $Q/J$
for small  values of frustration parameter  $Q/J$.  (The plot  for the
true Coulombic potential is very similar.)  The slope of each straight
line corresponds  to the Coulombic energy  and the  intercept with the
y-axis   corresponds to the   short-range ferromagnetic energy.  For a
given frustration parameter  $Q/J$, the ground  state is given by  the
straight-line that has  the smallest   energy.   When $Q/J\to 0$,   the
ground state  consists of lamellar phases  whose period becomes larger
and larger and whose  range  of stability  decreases.  The  values  of
$(Q/J)$  corresponding    to   transitions  between   two   successive
ground-state  structures  are    obtained  by  solving  the   equation
$E(m)=E(m+1)$.   Therefore, when   $Q/J$   goes   to zero,    $m$   is
asymptotically given by
\begin{equation}\label{eq:10}
m\sim(Q/J)^{-1/3},
\end{equation}
a behavior that is analogous  to that observed  for lamellar phases of
diblock copolymer systems at  low temperature in the  so-called strong
segregation limit\cite{MB96}. Note   that  for a large width    $m$, the
difference  between Eq.~(\ref{eq:9}) and   the value obtained fro the
true Coulombic term  is only weakly varying  with $m$ and can be taken
as a constant, so that Eq.~(\ref{eq:10}) provides also a very accurate
estimate both    for  the sequence of  lamellar   phases  and for  the
transition values of $(Q/J)$ for the true Coulombic potential.

\begin{figure}
\begin{center}

\resizebox{8cm}{!}{\includegraphics{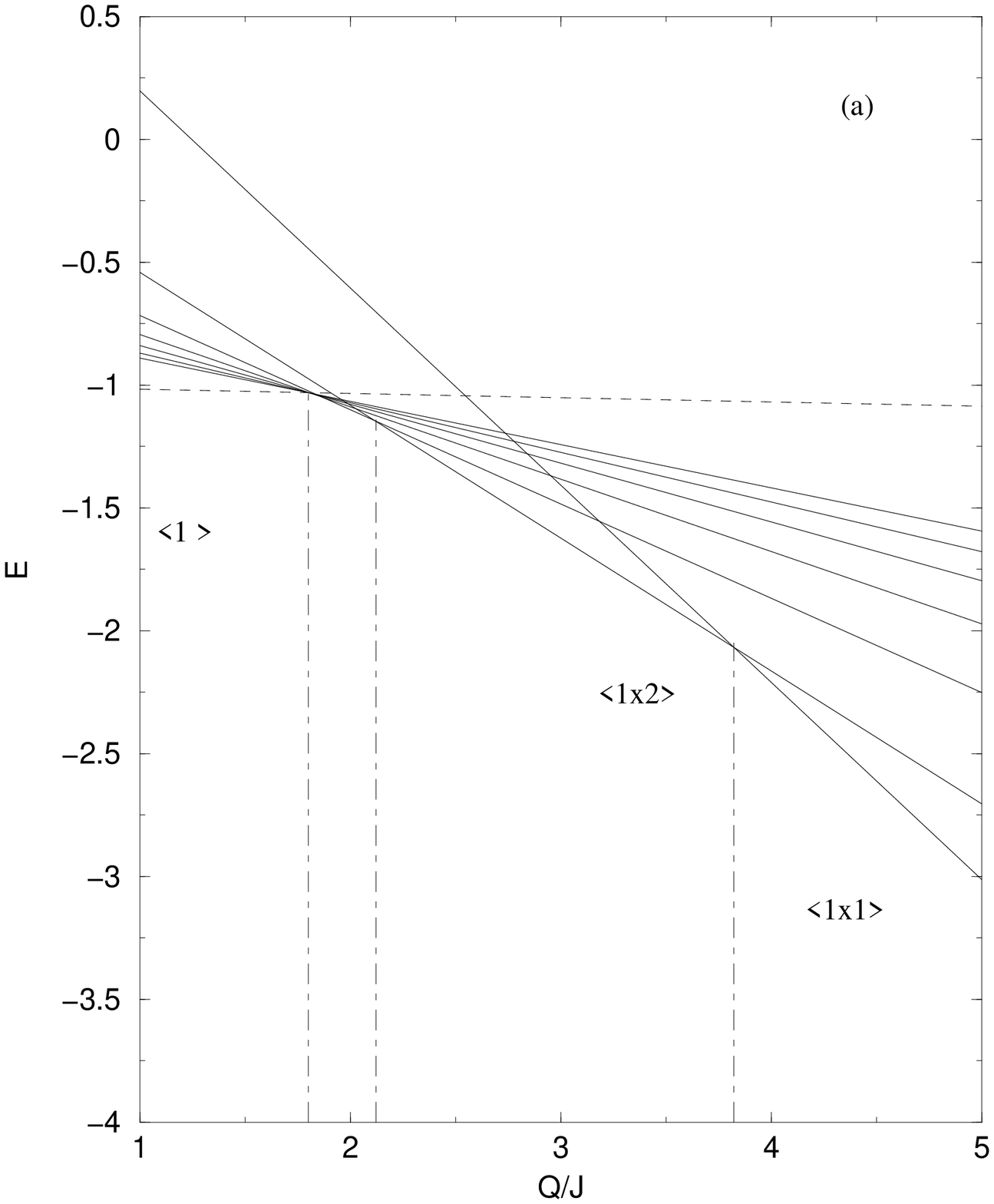}}
\resizebox{8cm}{!}{\includegraphics{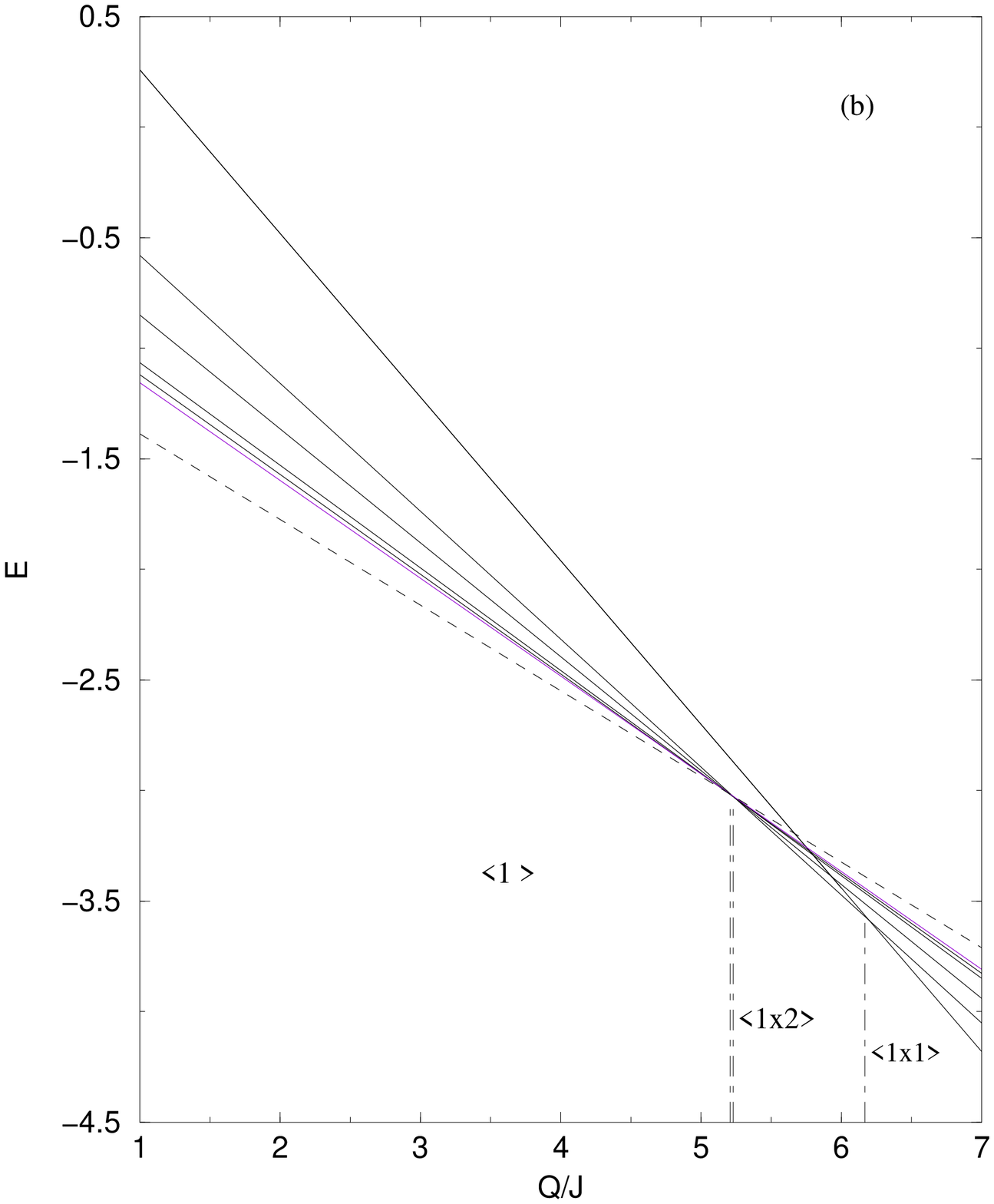}}
\caption{Ground-state energy of the tubular phases with $m=1,2,\ldots7$
(full lines) and $m=\infty$ (dashed line) as
a function  of the frustration  parameter  $Q/J$. (a) For the  inverse
lattice  Laplacian  expression, the tubular  phases,  $(1\times m_2 \times \infty )$
with $m_2>1$,, are  stable for frustration parameters  $1.800<Q<3.820$
(b) For  the true Coulombic  potential, the  tubular phases are stable
for $5.21<Q<6.17$. $<1>$, $<1\times 2>$ and $<1\times 1>$ are  short-hand
notations for $<1\times \infty\times \infty >$,  $<1\times 2\times \infty>$  and $<1\times 1\times \infty>$,
respectively. The vertical dot-dashed lines are visual guidances for
denoting the stability region of phases.}\label{fig:2}
\end{center}
\end{figure}

\begin{figure}
\begin{center}

\resizebox{8cm}{!}{\includegraphics{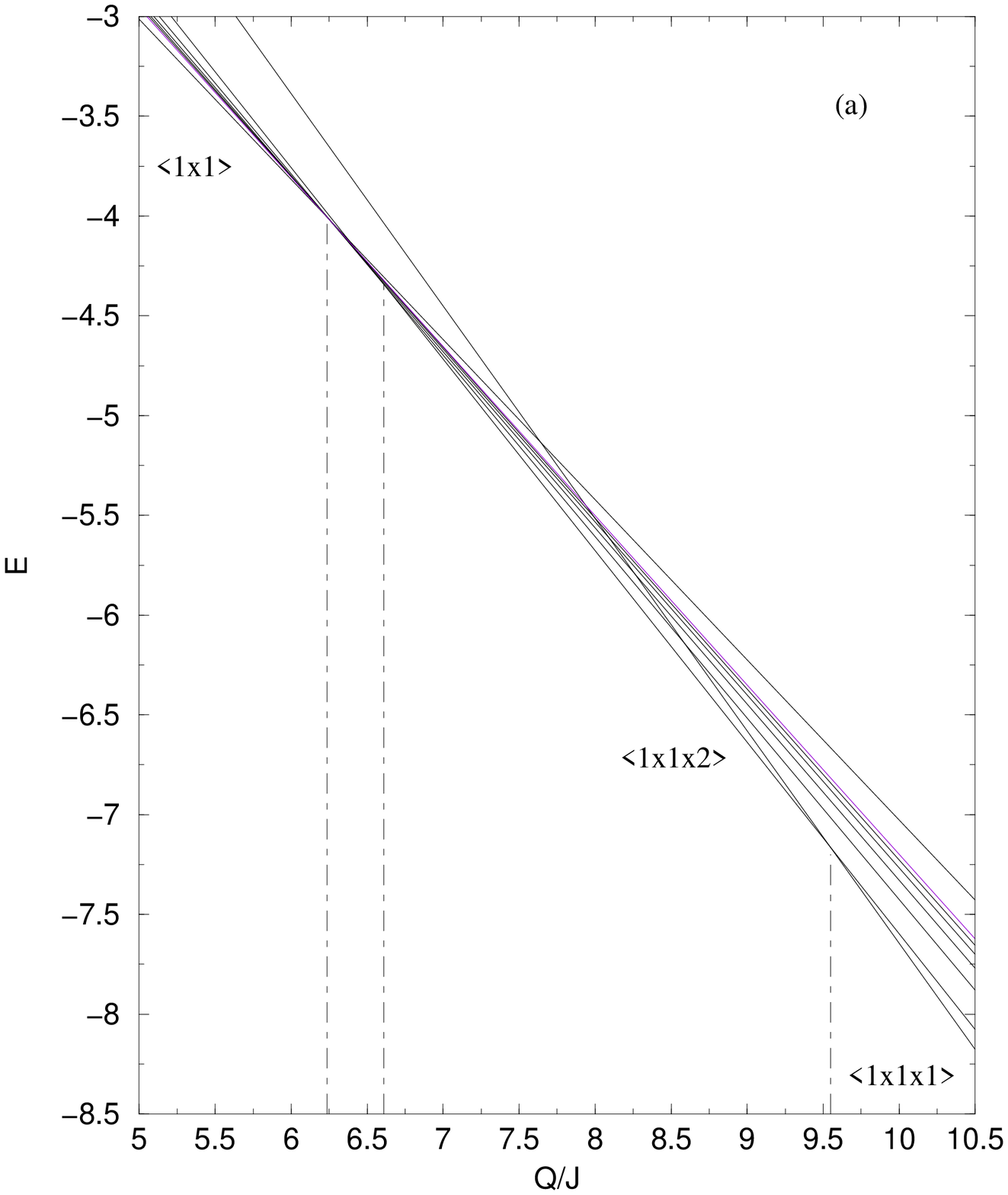}}
\resizebox{8cm}{!}{\includegraphics{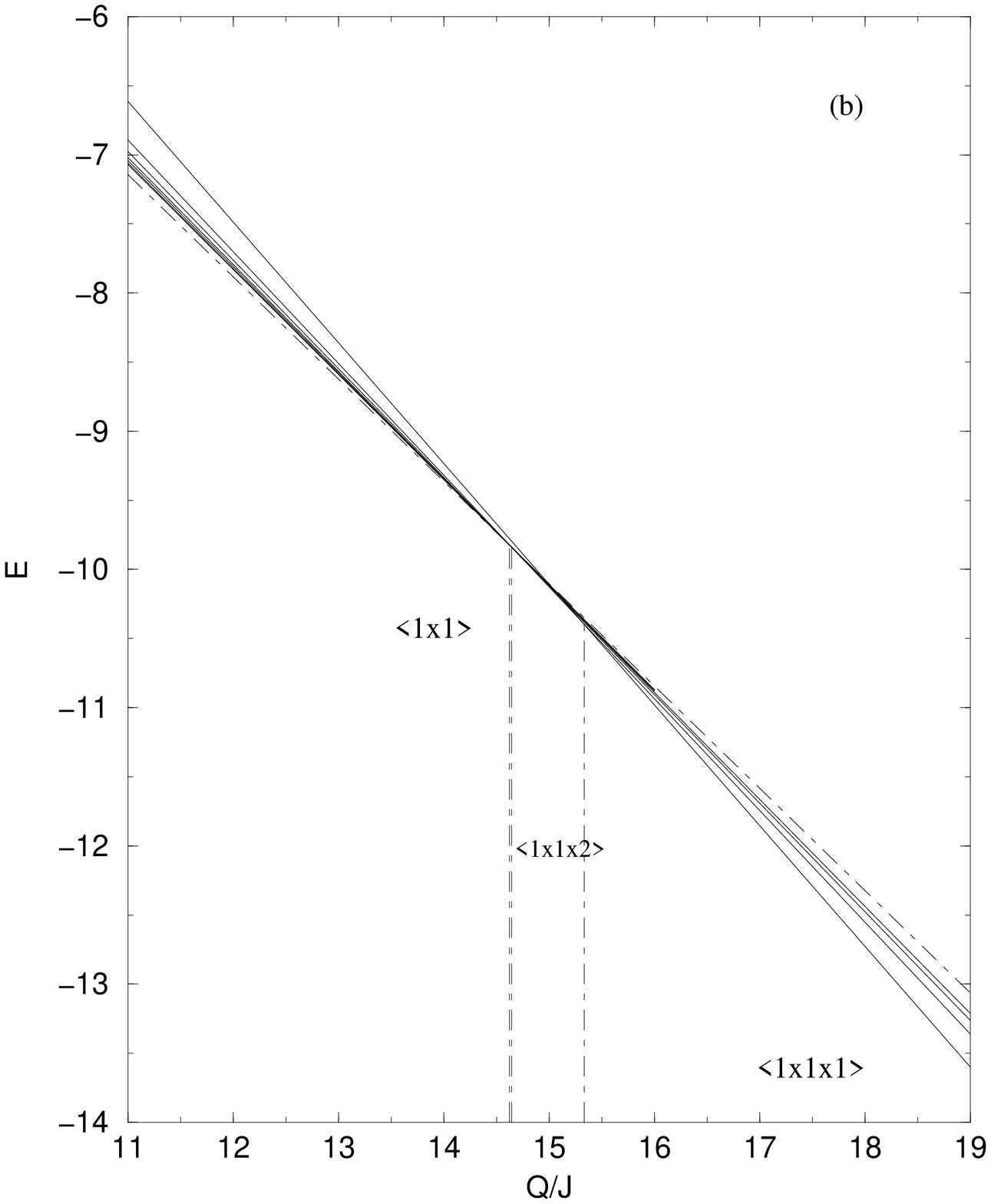}}
\caption{Energy of the orthorhombic phases with $m=1,2,\ldots7$ as a
function  of  the frustration  parameter  $Q/J$.  (a)  For the inverse
lattice Laplacian expression, the orthorhombic phases $(1\times  m_2 \times \infty )$
with $m_2>1$, are stable   for frustration parameters  $6.237<Q<9.549$
(b) For the true Coulombic potential,  the orthorhombic phases are for
$14.63<Q/J=15.33$.  $<1\times1>$   is a  short-hand
notations for $<1\times1\times \infty>$. The vertical dot-dashed lines are visual guidances for
denoting the stability region of phases. }\label{fig:3}
\end{center}
\end{figure}

When $Q/J$  increases, the period   of the  lamellar phases  decreases
until one   reaches $m=1$  for  $Q/J=0.637$.   For  a region   of  the
frustration parameter between $0.637$ and $1.800$, this lamellar phase
is then the most stable phase.  In a narrow interval  between
$Q/J=1.800$ and $Q/J= 2.122$ there is a  cascade of phases $(1\times m_2\times \infty
)$  with $m_2$ decreasing  until $m_2=2$ as  the frustration parameter
increases     (see   Fig.~\ref{fig:2}a).  For
$2.122<Q/J<3.820$, the stable  phase is  $(1\times  2\times  \infty )$.   Note  that
tubular phases of type $(m_1\times m_2\times  \infty )$ with  both $m_1$ and $m_2>1$
are never stable.  For $3.820<Q/J<6.237$, the ground state is a $1\times 1\times
\infty  $ tube.  In  another interval  $6.237<Q/J<6.611$, the system looses
translational invariance in the   third direction, and one  observes a
cascade of ``parallepipedic''   phases  $(1\times 1\times  m_3 )$,  with   $m_3$
decreasing as the frustration   increases until $m_3=2$.  Between
 $6.611$ and $9.549$, the stable  phase is  $(1 \times  1\times 2 )$.  For
$Q/J>9.549$, the ground state is the standard N{\'e}el  state $(1\times 1\times 1 )$
(see Fig.~\ref{fig:3}a).  It is noticeable that periodic structures of
the type $(m_1\times m_2\times m_3  )$ with $m_1,m_2,m_3>1$  and at least of the
$m_\alpha$'s  finite always  have higher   energies   than  those of   the
sequences $(m_1\times \infty  \times \infty )$, $(\infty \times  m_2\times \infty )$,  $(\infty \times \infty \times m_3 )$,
whatever   the    value    of   the   frustration     parameter   (see
Appendix~\ref{sec:coul-energy-simple}).

The same analysis can be repeated with $v({\bf r})$  given by the true
Coulomb   interaction.  The exact same    sequence of ground states as
before is obtained, but the values of the frustration parameter at the
transition points are  somewhat  shifted: For instance,  the  lamellar
phase  $m=1$ is  the most  stable   phase when $0.627<Q/J<5.21$,   the
cascade of  ``tubular''   phases  $(1\times  m_2\times   \infty  )$  occurs   around
$Q/J=5.22$, and $(1\times  2\times  \infty )$   is stable  for  $5.23<Q/J<6.17$ (see
Fig.~\ref{fig:2}b). The cascade of ``orthorhombic'' phases $(1\times 1\times
m_3  )$,    with  $m_3>1$, appears      around  $Q/J=14.63$ (See   Fig.
~\ref{fig:3}b).  The standard  N{\'e}el state  is stable for  $Q/J>15.33$.
It is worth mentioning that the counterparts of the lamellar, tubular,
and  orthorhombic phases  in diblock copolymer  systems\cite{MB96}
(systems that  are described at a coarse-grained  level by  the scalar
field theory associated with  the Hamiltonian in Eq.~(\ref{eq:1})) are
the lamellar, columnar, and cubic phases, respectively.

In  the  region of stability   of the lamellar   phases, we  have also
investigated     if    more  complex    structures  involving  several
one-dimensional modulations  of ferromagnetically ordered layers could
be present   and if multiphase transition  points  at which  more than
simply two phases coexist could occur.  For both questions, the answer
is negative.  As  an example, we give  in Appendix  ~\ref{sec:mix} the
analytical  expressions  for the energy  of  the mixed lamellar phases
that  are formed by  mixing  lamellae of width   $m=1$ and lamellae of
width $m=2$ with some given periodic modulation.  The simplest of such
phases, that following the notation used by Fisher  and Selke in their
study    of    the   axial   next-nearest-neighbor    Ising    (ANNNI)
model\cite{SF79}    we  denote  $<1^n2^p>$, consists    of  a periodic
repetition   of a fundamental  pattern formed  by  a succession of $n$
lamellae of width  $1$ followed by $p$ lamellae  of  width $2$.  (When
$n=0$,  one recovers the simple lamellar   phase of width $2$, denoted
$<2>$, and when $p=0$, one recovers the simple lamellar phase of width
$1$,  denoted $<1>$).  It is  easy  to check  that such mixed lamellar
phases are never stable at zero temperature.  In particular, they have
a  (strictly)  higher  energy than the  pure  lamellar  phases at  the
zero-temperature transition  point  between  the $<1>$ and   the $<2>$
phases,  at $Q/J=2/\pi$,   so that  this latter  is  a simple two-phase
coexistence  point.    This conclusion  remains    unchanged when  one
considers  even  more   complex  mixed   lamellar  phases,  such    as
$<1^{n_1}2^{p_1}  1^{n_2}2^{p_2}\ldots1^{n_s}2^{p_s}>$, with $n_\alpha ,p_\alpha $
integers   for $\alpha =1,2,\ldots s$,  whose  fundamental period is formed by
$n_1$ $1-$ layer lamellae followed by $p_1$  $2-$ layer lamellae, then
by $n_2$ $1-$ layer lamellae, and by $p_2$ $2-$ layer lamellae, etc\ldots,
two successive lamellae being composed of spins  of opposite signs. We
have also studied  one-dimensional  quasi-crystalline  arrangements of
lamellae: by  using binary   substitution rules,\cite{LGJJ93}  we have
built quasi-periodic structures by iteration, but we have always found
(numerically)  that   their energy is higher  than   that of  the pure
lamellar phases.  Note finally that from the calculation of the energy
of the mixed  lamellar phases at  zero temperature, one can also study
the change of  energy induced by adding  defects in the  pure lamellar
phase $<2>$.   As shown in Appendix ~\ref{sec:energy-defect-2}, adding
defects always cost energy, the  dominant effect being the increase in
the  short-range  energy.   The  same  analysis,   leading  to similar
conclusion, can  be repeated for the whole  range $0<Q/J<0.637  $ over
which lamellar phases are favored.

\section{Mean-field theory}\label{sec:mean-field-theory}
To  describe phases with a spatial  modulation  of the magnetization, we
consider a local mean-field approximation. If $m_i= <S_i>$ denotes the
local magnetization at site $i$ and  $\hat{m}(k)$ its lattice Fourier
transform, the mean-field free energy $ F_{mf}$ is given by:
\begin{equation}
\label{eq:11} \frac{\beta F_{mf}}{N}=-\frac{\beta J}{2N}\sum_{{\bf k}\neq 0}
\hat{V}({\bf k}) |\hat{m}({\bf k})|^2
-\frac{1}{N}\sum_{i}\ln(2\cosh(\beta H_{i})),
\end{equation}
where $\beta=1/k_BT$ and the effective field on site $i$, $H_i$, is equal
to $-J\sum_{j\neq i}V_{ij}m_j$, or in Fourier transformed space, 
\begin{equation}\label{eq:12}
\hat{H}({\bf k})=-J\hat{V}({\bf k})\hat{m}({\bf k})
\end{equation}

Minimization   of the free  energy  with  respect to the local magnetizations
 leads to the  self-consistent set of  equations: 
\begin{equation}
\label{eq:13}
m_i=\tanh(\beta H_i),
\end{equation}
for each  lattice site. One  must  then  solve the coupled  equations,
Eqs.(\ref{eq:12})   and (\ref{eq:13}) simultaneously, and subsequently
insert   the  solution  in  the   expression of  the   free  energy,
Eq.~(\ref{eq:11}). One  finally searches for  the configuration of the
$m_i$'s that  leads to the  deepest minimum of the  free  energy for a
given temperature and a given value of the frustration parameter.

To simplify the notation in the rest of the paper, units of
temperature, energy, etc\ldots, will be chosen such that $k_B=J=1$.

\subsection{Order-disorder transition line}\label{sec:order-disord-trans}

Close to   the transition between  the disordered  and the ordered phases, the
magnetizations $m_i$ are small and Eq.~(\ref{eq:13}) can be linearized,
\begin{equation}\label{eq:14}
m_i\approx   \frac{H_i}{T}.
\end{equation}
For a  given value     of  the frustration parameter,  the    critical
temperature $T_c(Q)$ is then given by
\begin{equation}\label{eq:15}
\label{eq:16}
T_c(Q)=-{\rm min}_{{\bf k}}\hat{V}({\bf k})=-V_c({\bf Q}),
\end{equation}
where  the minimum of $\hat{V}({\bf k})$,  $V_c({\bf Q})$, is attained
for a set of nonzero wave-vectors $\{{\bf k}_c(Q)\}$ that characterize
the   ordering   at $T_c(Q)$.   For   the   inverse  lattice Laplacian
expression of  the   long-range  frustrating  interaction, the   ${\bf
k}_c(Q)'$s vary continuously with $Q$ as follows
\begin{eqnarray}
{\bf k}_c = (\pm \arccos(1-\sqrt{\pi Q}), 0, 0),&\mbox{\hspace{1cm}} &
\mbox{\rm for $0\leq Q< 4/ \pi, $}\label{eq:17}\\
{\bf k}_c = ( \pi ,\pm \arccos(3-\sqrt{\pi Q}), 0),&\mbox{\hspace{1cm}} &
\mbox{\rm for $4/ \pi \leq Q< 16/ \pi, $}\label{eq:18}\\
{\bf k}_c = ( \pi , \pi ,\pm \arccos(5-\sqrt{\pi Q})),
&\mbox{\hspace{1cm}} & \mbox{\rm for $ 16/ \pi \leq  Q<36/ \pi, $}\label{eq:19}\\
{\bf k}_c = ( \pi , \pi , \pi),&\mbox{\hspace{1cm}} &\mbox{\rm for $36/ \pi \leq Q.$}\label{eq:20}
\end{eqnarray}
One should of course add all vectors obtained by permuting the $x,y,z$
coordinates  in Eqs.~(\ref{eq:17})-(\ref{eq:20}).   The above ordering
wave-vectors correspond,   respectively,   to lamellar (Eq.~(\ref{eq:17})),
tubular  (Eq.~(\ref{eq:18}), orthorhombic (Eq.~(\ref{eq:19})), and
cubic or  N{\'e}el (Eq.~(\ref{eq:20})) phases. The  corresponding critical
temperature $T_c(Q)$ is  then given by  (recall that both $T$ and  $Q$
are expressed in units of $J$)

\begin{eqnarray}
T_c(Q)=6-4\sqrt{\pi  Q}+4\pi Qv(0)&\mbox{\hspace{2cm}}& \mbox{\rm for
$0\leq Q \leq 36/ \pi $}\\
T_c(Q)=-6+4\pi Q\left(v(0)-\frac{1}{12}\right)&\mbox{\hspace{2cm}}& \mbox{\rm for
$36/ \pi\leq Q   $} 
\end{eqnarray}
The  mean-field approximation  gives  a  line  of second-order   phase
transition from the disordered to  the modulated phases, with $T_c(Q)$
first decreasing with  $Q$,   reaching a minimum   for $Q_{min}=1/(4\pi
v(0)^2)=1.245871$ at  $T_c(Q_{min})=6-1/v(0)$ and  then
increasing again for finally reaching a regime of linear increase with
$Q$ when $Q\geq 36/ \pi $.  It is worth noting  than the term $4\pi Qv(0)$
is  important  for  large  frustration:  indeed,  since $v(0)>1/6$, it
allows  to obtain a positive  critical  temperature for all $Q'$s. (In
general, $v(0)$ should be larger than the inverse of the number of nearest
neighbors on the lattice, e.g..  $6$ on a simple  cubic
lattice).    For  vanishing  small
frustrations,   the   critical  temperature    goes   continuously  to
$T_{c}^{0}$, the critical  temperature of the pure Ising  ferromagnet.

\subsection{Structure combination branching processes}\label{sec:full-phase-diagram}

\begin{figure}
\begin{center}
\resizebox{8cm}{!}{\includegraphics{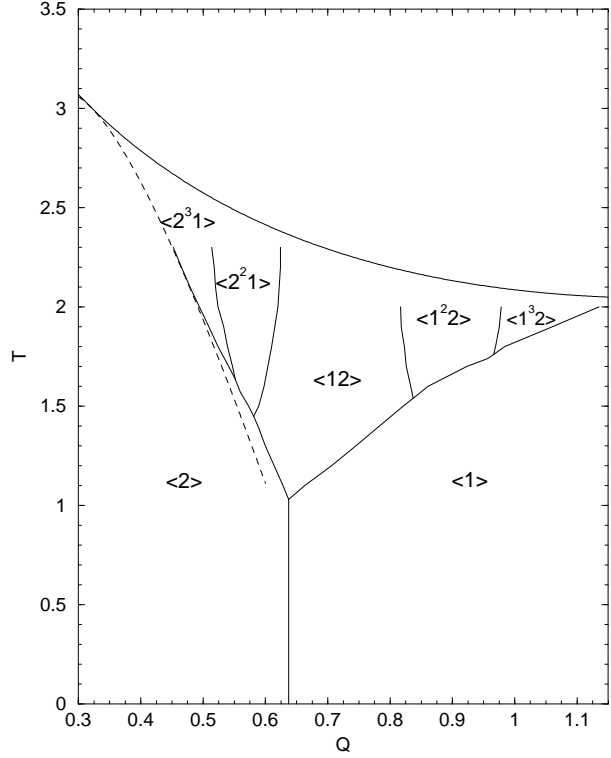}}
\caption{Mean-field phase diagram: structure combination  branching processes
occurring at finite temperatures for a region of frustration parameter
where the  ground  states consists of  $<1>$ and  $<2>$   phases.  The
dashed line corresponds  to the  soliton-approach  prediction for  the
(upper) stability of  the $<2>$ phase. The  units are chosen such that
$k_B=J=1$.}\label{fig:4}
\end{center}
\end{figure}

At zero  temperature, we    showed  that the  system  exists  in  pure
modulated phases, whose modulation is commensurate with the underlying
lattice.   At  the transition  between  the  modulated and  disordered
phases,  we have   just   seen that the  ordering  wave-vector  varies
continuously with  the frustration parameter,  hence indicating that a
succession of incommensurate modulated phases   is observed along  the
critical  line.  As the temperature increases  from $T=0$ to $T_c(Q)$,
one expects a  cascade of  ordered  phases with  commensurate  spatial
modulations of increasing complexity until a point is reached at which
incommensurate phases appears; this  is what is observed  for instance
in  the    much studied  axial   next-nearest neighbor   Ising (ANNNI)
model\cite{S88},  in which  the    cascade of phases  is   produced by
``structure combination   branching  processes''\cite{S88}.   To
illustrate how such branching processes proceed, we consider first the
low-temperature region   of the phase diagram   in which  the simplest
modulated  phases, the  $<1>$ and  $<2>$ lamellar  phases, are stable  (see
Fig. \ref{fig:4}).   Eqs.~(\ref{eq:12}) and (\ref{eq:13})  must now be
solved beyond  the linear approximation.   The $<1>$ phase corresponds
to  an alternate configuration  of layers of  up and  down spins, with
$m_i=m_1(-1)^i$, which implies that the wave-vector characterizing the
modulation has only   one nonzero component equal  to   $k=\pi$.  Using
Eq.~(\ref{eq:12}) and Eq.~(\ref{eq:13}), one gets
\begin{equation}
\label{eq:22} 
m_1=\tanh\left[ \frac{m_1}{T}(2-\pi Q+4\pi Q v(0))\right],
\end{equation}
which has a nonzero solution when
\begin{equation}
\label{eq:23} 
 T<2-\pi Q+4\pi Qv(0).
\end{equation}
From Eq. (\ref{eq:11}),  the corresponding free energies of the
two phases is obtained as 
\begin{equation}
\label{eq:24}
\frac{F_{<1>}}{N}=\left(1-\frac{\pi Q}{2}+
2\pi Qv(0)\right)m_{1}^{2}-T\ln\left[2\cosh\left(\frac{m_1}{T}(2-\pi Q+4\pi Qv(0))\right)\right].
\end{equation}

The $<2>$  phase corresponds to   an alternate  sequence  of pairs  of
ferromagnetically ordered layers and is characterized by a wave-vector
whose  only  nonzero component is $k=\frac{\pi}{2}$.  The corresponding
order parameter $m_2$ satisfies the following equation:
\begin{equation}
\label{eq:122} 
\frac{\sqrt{2}}{2}m_2=\tanh\left[ \frac{\sqrt{2}m_2}{T}(2-\pi Q+2\pi Q v(0))\right],
\end{equation}
which has a nonzero solution when 
\begin{equation}
\label{eq:123} 
 T<2-\pi Q+2\pi Qv(0).
\end{equation}
The associated free energy is 
\begin{equation}
\label{eq:124}
\frac{F_{<2>}}{N}=\left(1-\frac{\pi Q}{2}+
\pi Qv(0)\right)m_2^2-T\ln\left[2\cosh\left(\frac{m_2\sqrt{2}}{T}(2-\pi Q+2\pi Qv(0))\right)\right].
\end{equation}
The line of first-order transition at  which the two phases coexist is
defined by $F_{<1>}=F_{<2>}$. In the $T-Q$ phase diagram, this line is
almost vertical with $Q\simeq  2/ \pi $. As  we  have already stressed,  no
mixed phases  coexist with  the $<1>$  and  $<2>$ phases at  $T=0$ and
$Q=2/ \pi $. However, the mixed $<12>$ phase  may become more stable at
a nonzero temperature. This phase  is characterized by the  modulation
$m_i=\frac{2\sqrt{3}}{3} m_{12} \cos(2\pi  i/3+\pi /2)$, with  the order
parameter $m_{12}$ determined through the self-consistent equation
\begin{equation}
\label{eq:222} 
m_{12}=\tanh\left[ \frac{m_{12}}{T}\left(3- \frac{4\pi}{3} Q+4\pi Q v(0)\right)\right],
\end{equation}
which has a nonzero solution for 
\begin{equation}
\label{eq:223} 
 T<3- \frac{4\pi}{3} Q+4\pi Q v(0).
\end{equation}
The corresponding free energy is given by 
\begin{equation}
\label{eq:324}
\frac{F_{<12>}}{N}=\frac{1}{3}\left(3-\frac{4\pi Q}{3}-
\pi Qv(0)\right)m_{12}^2-\frac{2T}{3}\ln\left[2\cosh
\left(\frac{m_{12}}{T}(3-\frac{4\pi}{3}Q+4\pi Qv(0))\right)\right].
\end{equation}
 
Comparing now  the free  energy  of   the three phases,   $<1>$,$<2>$,
$<12>$, one can show that the $<12>$ phase become more stable than the
other two  in   a wedge above   a  branching point at  $T_b=1.03$  and
$Q=0.63$, at which the three phases coexist. This represents the first
step   of  a structure  combination    branching process  by which two
adjacent phases,  here $<1>$  and  $<2>$, get separated  above a given
branching point at finite temperature by a  phase corresponding to the
simplest combination structure,   here  the $<12>$ phase.   A  careful
examination of the thermodynamic quantities  shows that the entropy of
the $<12>$ phase increases more rapidly with  temperature than that of
the $<1>$ and   $<2>$ phases; although  it has   an unfavorable energy
contribution,  the $<12>$ phase  becomes thermodynamically stable when
the temperature becomes  high enough ($T_b=1.03$).  Moreover,  for all
temperatures between $T=0$ and $T_b$, both the  entropy and the energy
of the $<1>$ and $<2>$ phases are identical along the coexistence line
between the two phases.

To  study in more  detail the branching processes,   one can no longer
rely  on explicit analytical calculations  as done above, because they
become rapidly intractable.   To obtain  the phase  diagram  in a more
systematic        way       ,      the      mean-field      equations,
Eqs.~(\ref{eq:11})-~(\ref{eq:13}),   can be  solved iteratively    for
finite    lattices    with    periodic  boundary    conditions.    The
thermodynamically stable    solution  corresponds to    that  with the
smallest free  energy.  For each  temperature  it is  assumed that the
spin structure repeats  itself  after  $N$ layers  (only  commensurate
lamellar   structures   are considered).    The    iterative procedure
converges whenever the  initial configuration is  not too far from the
equilibrium one.  The iterative sequence  is as follows. The effective
fields $H_i$ are calculated from the set of initial magnetizations $\{
m_i^{0}\} $ via Eq.~(\ref{eq:13}).  One computes the Fourier transform
of the fields $H_i$, and using Eq.~(\ref{eq:12}) yields then a new set
of magnetizations $\{ m_i^{1}\}  $ which is used as  the input for the
next iteration.     The calculation should  be  performed  for various
values of  $N$ for examining  many different  commensurate structures.
We have carried out the  calculation for N up  to 16.  The phases with
large-N periodicities, like in the  ANNNI phase diagram\cite{S88}, are
only stable in a small neighborhood of the  critical line.  Therefore,
the most  of  the phase  diagram, except  in the  region  close to the
critical   line,  can be  drawn    by considering simple  commensurate
wave-vectors.

The  results  are shown in   Figs.~\ref{fig:4} and \ref{fig:5}. Figure
\ref{fig:4}   illustrates  the structure  combination   process in the
region between the $<1>$ and $<2>$ phases. The $<12>$ and $<1>$ phases
are separated  by  the  $<1^22>$ phase which  is  itself  separated at
higher temperature from the $<1>$ phase by the  $<1^32>$ phase, and so
on, until  one presumably   reaches   an accumulation point  of    the
branching  process along the   transition line  above which the  $<1>$
phase melts.    Beyond  this accumulation   point corresponding   to a
sequence of  $<1^n2>$ phases when $n\to   \infty $, devil's  staircases and
incommensurate phases  are    expected\cite{S88}.   Figure~\ref{fig:5}
provides a broader picture  of the phase diagram  in the region  
where the modulated phases are lamellar:  the phase diagram appears as
a succession of flowers of complex modulated phases, separated by regions
in which the pure lamellar phases $<1>$,  $<2>$, etc\ldots, are stable; the
flowers get closer and closer when  the frustration decreases. Note
that  the  first branching  point  at which   the simplest mixed phase
appears is always at a non-zero temperature.

\subsection{Soliton approach}
\label{sec:Soliton-theory-of-the-modulated-phase}

At high enough temperature, near the critical line, the modulated
phases are incommensurate since we saw in
section~\ref{sec:order-disord-trans} that the ordering wave-vector
varies continuously with frustration. More insight in the phase behavior
can be then provided by employing the soliton approach developed by
Bak and coworkers\cite{PB80,JB84}. More precisely, this approach allows one
to study analytically the melting of a commensurate phase to
incommensurate phases by focusing on the behavior of the domain walls
that separate commensurate regions, domain walls that can be
considered as ``solitons''. In the following, we use the soliton method to
investigate the stability of commensurate phases at high temperature.

In the vicinity of the (upper) melting line of the $<2>$ phase, one
can expand the mean-field equations in the appropriate order
variables, which leads to the following free-energy functional

\begin{eqnarray}
\label{eq:30}
 \frac{F}{N}=-\frac{1}{2}\sum_{{\bf k}}[\hat{V}({\bf k})+\frac{1}{\beta} ]|m_{{\bf k}}|^2+
\frac{T}{12}\sum_{{\boldsymbol \tau}}\sum_{{\bf k_1}\ldots{\bf k_4}}
\delta({\bf k_1}+{\bf  k_2}+  {\bf k_3}+{\bf k_4}-{\boldsymbol  \tau}) \hat{m}_{{\bf k_1}}
\hat{m}_{{\bf        k_2}}\hat{m}_{{\bf        k_3}}\hat{m}_{{\bf
k_4}}\nonumber\\
+\frac{T}{30}\sum_{{\boldsymbol     \tau }}\sum_{{\bf  k_1}\ldots{\bf k_6}} 
\delta({\bf  k_1}+{\bf k_2}+ {\bf k_3}+{\bf k_4}+{\bf
k_5}+{\bf k_6}-{\boldsymbol \tau})  \hat{m}_{{\bf
k_1}} \hat{m}_{{\bf k_2}}\hat{m}_{{\bf k_3}}\hat{m}_{{\bf k_4}}\hat{m}_{{\bf k_5}}
\hat{m}_{{\bf k_6}}+\ldots
\end{eqnarray}

where a   constant   term has been discarded   and   ${\boldsymbol \tau}$  is  a
reciprocal-lattice  vector. The  above  expression, Eq.~(\ref{eq:30}),
contains both regular and ``umklapp''   terms; these
latter, represented by the second and third contributions in the
right-hand side of  Eq.~(\ref{eq:30}), correspond to terms in which
the sum of the wave-vectors is equal to a reciprocal-lattice vector,
i.e., they keep track of the underlying lattice structure and are
responsible for the stability of the commensurate phases.

For studying the stability of the  $<2>$ phase near the critical line,
we consider  wave vectors that  are close to  the ordering wave vector
$k_{\pi/2}=(\pi /2,0,0)$ with small fluctuations  $q_x$ in the direction
of  the modulation, here    along the $x-$axis,    and ${\bf q}_\bot$  in  the
perpendicular  layer. For   the  present  case,  it is  sufficient  to
truncate Eq.~(\ref{eq:30})   after  the fourth  order\cite{PB80},  and
after  expanding    the  interaction term    to  second  order  in the
fluctuations the free-energy functional can be rewritten as

\begin{eqnarray}
\label{eq:31}
\frac{F}{N }=\sum_{{\bf q}=q_x,{\bf q}_\bot,}\left(r+ak_x+cq_{x}^{2}+c'{\bf
q}_{\bot}^{2}\right)|m_{\bf
k_{\pi/2}+q }|^2  + T \sum_{\bf q_1 q_2 q_3 q_4}\delta({\bf
k_1+k_2+k_3+k_4})\nonumber\\
\left[\frac{1}{2}m_{\bf k_{\pi/2}+q_1}m_{\bf k_{\pi/2}+q_2} m_{-\bf
k_{\pi/2}+q_3} m_{-\bf k_{\pi/2}+q_4}
+\frac{1}{12}(m_{{\bf k_{\pi/2}+q_1}}m_{{\bf k_{\pi/2}+q_2}}
m_{{\bf k_{\pi/2}+q_3}}m_{{\bf k_{\pi/2}+q_4}}\right.\nonumber \\\left.
+m_{{-\bf k_{\pi/2}+q_1}}m_{{-\bf k_{\pi/2}+q_2}}
m_{-{\bf k_{\pi/2}+q_3}}m_{-{\bf k_{\pi/2}+q_4}})\right],
\end{eqnarray}
where $r=T-4+2\pi Q-4\pi Qv(0)\leq 0$, $a=2-2\pi Q$, $c=2\pi Q$, and
$c'=-\pi Q+1$. Introducing now two continuous order parameters, 
\begin{eqnarray}
\label{eq:32}
m_{+}({\bf r})&=&\sqrt{2}\int \frac{d^3q}{(2\pi)^3}\exp(i{\bf qr})
m_{{\bf k_{\pi/2}+q}}\nonumber\\
m_{-}({\bf
r})&=&\sqrt{2}\int\frac{d^3q}{(2\pi)^3}\exp(i{\bf
qr})m_{{\bf -k_{\pi/2}+q}}, 
\end{eqnarray}
where $m_{-}({\bf r})$ is the complex conjugate of  $m_{+}({\bf r})$,
one can express the free-energy functional in the following
Ginzburg-Landau form: 
\begin{eqnarray}
\label{eq:33}
\frac{F}{N}&=&\int d^3{\bf r}\left(\frac{1}{2}c\left|\left(\frac{\partial }{\partial
x}-i\frac{a}{2c}\right)m_+\right|^2+\frac{1}{2}c'|{\bf  \nabla}_\bot  m_+|^2
+\frac{1}{2}\left(r-\frac{a^2}{4c}\right)|m_+|^{2}\right.\nonumber\\
&+&\left.\frac{T}{8}|m_+|^{4}+\frac{T}{8}(m_+^{4}+m_-^{4})\right),
\end{eqnarray}
where     ${\bf \nabla}_\bot=(0,\frac{\partial }{\partial y}, \frac{\partial }{\partial z}) $ and the
last term is generated by the umklapp terms. Following Bak and
Boehm\cite{PB80}, we choose the following ansatz for the order
parameters,
\begin{equation}
\label{eq:35}
m_{\pm}=A\exp(\pm i\phi(x)),
\end{equation}
where the  amplitude $A$ is a constant. Note that $\phi (x)$ is constant in
the commensurate $<2>$ phase and that $A$ can be obtained by
minimizing the free energy in that phase, which gives $A^2=3|r|/ T $.
Inserting the above expression in Eq.~(\ref{eq:33}) leads, up to a
constant, to the following free-energy functional per unit area
perpendicular to the $x-$direction:
\begin{equation}
\label{eq:36}
F[\phi]=\frac{cA^{2}}{2}\int dx \left[\phi^{'2}(x)-\frac{a}{c}\phi^{'}(x)-
\frac{TA^2}{12c}(1-\cos4\phi(x))\right],
\end{equation}
where the first term is minimized for $\phi(x)=Ax/(2c)$, which
corresponds to an incommensurate spatial modulation, whereas the
second term is minimized for $\phi(x)=0$, i.e., in the commensurate
$<2>$ phase. The overall minimization of $F[\phi]$ is attained when the
phase function $\phi(x)$ obeys  the sine-Gordon equation:
\begin{equation}
\label{eq:37}
\phi''(x)+4v\sin(4\phi(x))=0,
\end{equation}
where $v=TA^2/(24c)$.   The  solution consists of  regions of constant
phase separated by solitons in  which $\phi$ increases by $\frac{\pi}{2}$
over a  short distance.  More generally,  one can lock for a solution
over   a (large) distance  $L$  that  consists  of $n$  equally spaced
solitons where $ \frac{n\pi }{2}=L\phi'$. 
The corresponding free-energy is given by\cite{bak76,perring61} 
\begin{equation}
\label{eq:38}
\frac{F}{cA^{2}L}=\left[\frac{4}{\pi}v^{\frac{1}{2}}-
|\frac{-a}{2c}|\right]\phi'+\frac{16}{\pi}v^{\frac{1}{2}}\phi'
\exp[-\frac{2\pi}{\phi'}v^{\frac{1}{2}}].
\end{equation}
where $\phi'=n\pi  /(2L)$.  The first  term, proportional  to the soliton
density, is  the formation energy; the  second  term corresponds  to a
weak repulsion between solitons.  The commensurate phase is stable  as
long as  the first term remains positive;  otherwise, the  $<2>$ phase
becomes unstable  with the respect  to soliton formation. The critical
temperature corresponding to this melting transition is given by
\begin{equation}
\label{eq:39}
T_{<2>-IC}=4-2\pi Q+4\pi Qv(0)-\frac{\pi}{4}\frac{(1-\pi Q)^{2}}{Q}.
\end{equation}
and is shown as a dashed line in Figs.~\ref{fig:4} and \ref{fig:5}.

A similar analysis can be performed to study, for instance, the
stability of the $<3>$ phase near the critical line. The main changes
are that the relevant ordering wave-vector is now $(\pi /3,0,0)$ and
that the sixth-order umklapp terms should be kept in
Eq.~(\ref{eq:30}). A transition between the $<3>$ phase and
incommensurate phases is found for 
\begin{equation}
\label{eq:41}
T_{<3>-IC}=5-4\pi Q+4\pi Qv(0)-\frac{3\pi}{4}|1-4\pi Q|\sqrt{\frac{5(5-4\pi
Q)}{1+20\pi Q}}.
\end{equation}
The result is shown as dotted curve in Fig.~\ref{fig:5}. The  deviation from
the numerical solution of the  mean-field  equations that is seen  when
moving further away from the critical line comes from the truncation
of the free-energy functional that is used in the soliton approach.
\section{Discussion}\label{sec:discussion}
\begin{figure}
\begin{center}

\resizebox{8cm}{!}{\includegraphics{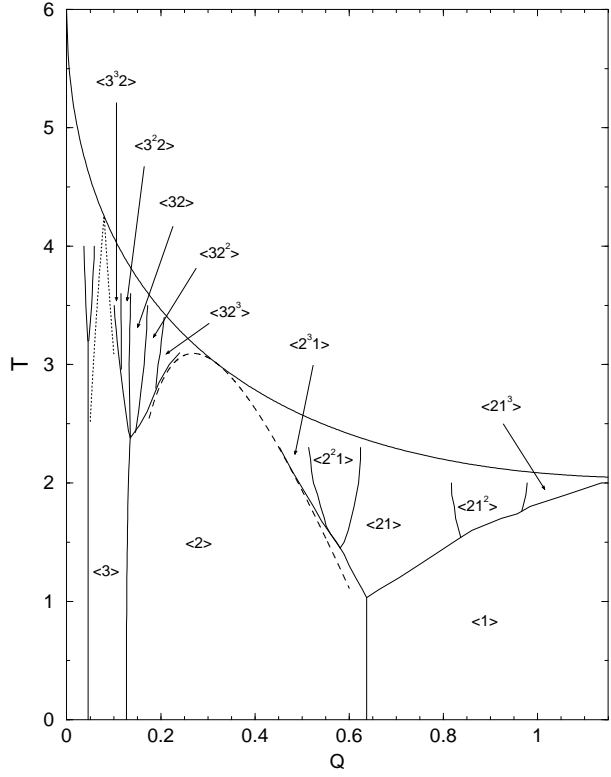}}
\caption{Mean-field phase diagram: a (partial) view of the infinite
sequence   of ``flowers'' of    complex modulated phases appearing  at
finite temperatures  for the range  of the frustration parameter where
the  ground states are  lamellar phases.  The  dashed and dotted lines
correspond  to    the soliton-approach  prediction   for  the  (upper)
stability of the $<2>$ and  $<3>$  phases, respectively. The units  are
chosen such that $k_B=J=1$.}\label{fig:5}
\end{center}
\end{figure}

The fact that spin models with competing interactions can give rise to
complex spatially  modulated   phases  has  been known   for   several
decades. The ANNNI model is one of  the simplest and best-studied such
system,  in  which  Ising spin   variables  situated on  a lattice are
coupled  via   nearest-neighbor    ferromagnetic  interactions      in
($D-1$)-dimensional layers  orthogonal to, say,  the $x-$ axis and via
next-nearest  neighbor antiferromagnetic  interactions along the  $x-$
axis\cite{S88}. The major additional ingredient that is present in the
Coulomb frustrated Ising ferromagnet and not is the previously studied
models   is the long-range  nature of  the competing antiferromagnetic
interaction.  It  is then   worth  reviewing some of  the  differences
between  the    phase  behavior  of    the  three-dimensional Coulomb
frustrated  ferromagnet and that   of the three-dimensional  ANNNI and
related  models\cite{UY89}.   First,  the   long $1/r$  range  of  the
frustrating interaction forbids ferromagnetic ordering for any nonzero
value of the frustration parameter $Q$, and, as  a result, there is no
Lifshitz point\cite{HLS75} in the model.  Secondly, there is no highly
degenerate multiphase point at  zero temperature; this is  in contrast
with   the ANNNI model  that     possesses an infinitely    degenerate
multiphase point  at zero temperature, a  point from which  springs an
infinite  number of distinct commensurate modulated phases\cite{DS83}.
Thirdly,   the  phase  diagram     of  the  Coulomb  frustrated  Ising
ferromagnet, as illustrated  in Fig.~\ref{fig:5}, displays an infinite
number of distinct flowers of  complex spatially modulated phases that
emerge from zero-temperature two-phase  coexistence points, the extent
of the flowers  in $T-Q$ phase  diagram decreasing as  the frustration
parameter decreases.

The long range  of  the antiferromagnetic interaction in  the  Coulomb
frustrated ferromagnet also brings about additional limitations on the
mean-field approach. On general grounds, one can expect the mean-field
description to reproduce the topology  of the phase diagram correctly,
but to become increasingly inaccurate as one approaches the transition
line, both from below and from above, because it overlooks the role of
fluctuations.   In the present case,    the fluctuations have a  major
effect  on the transition line: as  argued  by Brazovskii\cite{B75} on
the   basis  of   a    self-consistent   Hartree   treatment   of    a
field-theoretical  model analogous to   the Coulomb  frustrated  Ising
ferromagnet, and confirmed  by Monte Carlo simulations\cite{VT98}, the
fluctuations drive  the  order-disorder   transition from second    to
first-order.   The  mean-field approach is  thus   questionable in the
vicinity  of the transition line  (which  is why it   may not be worth
pursuing the search for  devil's staircases as was  done for the ANNNI
model\cite{PB80,BB82}). However, the  main  points reviewed above  are
not affected.



\newpage
\appendix

\section{Calculation of the Coulombic energy of lamellar phases in real
space} \label{sec:calc-coul-energy}

The   calculation  of the ground-state energy    due to the long-range
Coulombic   interaction can  be  performed  as  follows. For  lamellar
phases, the   sum over the reciprocal  vectors  is performed along one
direction.  One   then    obtains    the  one-dimensional    potential
corresponding to the inverse lattice Laplacian\cite{K89},
\begin{equation}\label{eq:42}
W(i)=2\pi Q \frac{\exp(-\alpha |i|)-1}{\alpha },
\end{equation}
where $\alpha $ is a convergence factor that will taken to zero at the end
of the  calculation. The  introduction  of $\alpha  >0$ in the calculation
allows   one to handle conditionally   convergent  sums.  The  average
Coulombic energy per site  of a cell  $2m$ is given by
\begin{equation}\label{eq:43}
E_C=\frac{1}{4m}\sum_{N'=-\infty }^{+\infty}\sum_{i,j}^{2m} W(|i-j+2mN'|)S_iS_j,
\end{equation}
where $N'$  is the index  for labelling the right  and  left cells and
$i,j$ are  denote sites within the cell.   The energy per  site can be
divided into two parts: the first one comes from the interaction between
the   spins within   the cell   and   the second one   comes  from the
interaction between a  spin and its images  in  the other  cells; this
reads
\begin{eqnarray}
E_C&=&E_1+E_2\nonumber\\                  &=&\frac{1}{4m}\sum_{i,j}^{2m}
W(|i-j|)S_iS_j+\frac{1}{4m}\sum_{N'\neq0}\sum_{i,j}^{2m}
W(|i-j+2mN')S_iS_j.\label{eq:44}
\end{eqnarray}
By considering  all contributions between pairs  of sites within the cell, one
gets for $E_1$
\begin{equation}\label{eq:45}
E_1=\frac{1}{2m}\left(
m(W(0)-W(m))+\sum_{n=1}^{m-1}(2(m-n)-n)W(n)-\sum_{n=m+1}^{2m-1}((2m-n)-n)W(n)\right).
\end{equation}
After some calculation, and taking the limit  $\alpha\to0$ at the end, one gets
\begin{equation}\label{eq:46}
E_1=Q\left(\frac{2\pi m^2 }{3}+\frac{\pi }{3}\right).
\end{equation} 

The sum over the  right and the  left cells can also be performed, and
$E_2$ is then given by 
\begin{equation}\label{eq:47}
E_2=\frac{\pi Q}{2m}\sum_{i,j}\frac{e^{-\alpha (2m+i-j)}+e^{-\alpha(2m+j-i)}}{\alpha(1-e^{-\alpha 2m})}S_iS_j.
\end{equation}
By using  the electro-neutrality  condition ($\sum_iS_i=0$), and in the limit $\alpha  \to
0$, one obtains $E_2$, which has a finite value:
\begin{eqnarray}
E_2&=&\frac{\pi Q }{4m^2}\sum_{i,j}(i-j)^2S_iS_j\nonumber\\
&=&-\frac{\pi Q }{2m^2}|\sum_{i=1}^{2m}iS_i|^2\nonumber\\
&=&-\frac{\pi Q m^2}{2}.\label{eq:48}
\end{eqnarray}
This gives for the Coulombic energy $E_c$
\begin{equation}\label{eq:49}
E_c=Q\left(\frac{\pi m^2}{6}+\frac{\pi }{3}\right).
\end{equation}
By subtracting the self-energy of the inverse lattice Laplacian potential
to Eq.~(\ref{eq:49}), one exactly recovers Eq.~(\ref{eq:9}).
 
\section{Ground-state energy of tubular and orthorhombic phases}
\label{sec:coul-energy-simple}
Let  us calculate the energy per  site for configurations whose phases
are  periodic  with  a orthorhombic  cell  ($m_1\times m_2\times
m_3$). The short-range energy per spin is obtained as
\begin{equation}\label{eq:21}
E_{SR}=-J\left(3-\frac{2}{m_1}-\frac{2}{m_2}-\frac{2}{m_3}\right)
\end{equation}
In the reciprocal space, the allowed wave-vectors have components $((2n_1+1)\pi /m_1,(2n_2+1)\pi /m_2,(2n_3+1)\pi /m_3)$, with $0\leq n_1 <  m_1-1$,
$0\leq n_2  <m_2-1$, $0\leq n_3 <m_3-1$,  so that the Coulombic energy per
spin $E_c$ is given by
\begin{equation}\label{eq:50}
E_c=Q\left[\sum_{n_1 =0}^{m_1-1}\sum_{n_2 =0}^{m_2-1}\sum_{n_3 =0}^{m_3-1}
\left(\frac{\pi  }{(\sum _{\alpha=1}^{3}\sin\left(\frac{(2n_\alpha+1)\pi}{2m_\alpha}\right)^2)
\prod_{\alpha=1}^{3}\sin\left(\frac{(2n_\alpha +1)\pi}{2m_\alpha }\right)^2} \right) -2\pi v(0)\right]
\end{equation}

When both $m_2$ and $m_3$ go to  infinity, one obtains lamellar phases
of period $2m$, and Eq.~(\ref{eq:50}) reduces to Eq.~(\ref{eq:9}).

When the periodic structure looses  translational invariance in a
second direction, one obtains a lattice of tubes whose
perpendicular section is a rectangular  cell $m_1\times m_2$.
Although we have not obtained a fully analytical
expression for such phases, some results can
be derived. The  Coulombic  energy for tubes of section $m_1 \times m_2$ is
bounded as follows:
\begin{equation}
 E_c(inf(m_1,m_2)) \leq    E_c(m_1,m_2)\leq E_c(sup(m_1,m_2)).
\end{equation}
If $p \times m_1=m_2=m$ where $p$ is a positive  integer, the energy $E_c(m)$
behaves as
\begin{equation}
E_c(m^2)\simeq C(p)m^2+ O(m),
\end{equation}   
where the numerical coefficients
$C(p)$ are summarized in table \ref{tab:1}.

We  have also calculated numerically, via Eqs.~(\ref{eq:21}) and
(\ref{eq:50}) the total
energy of modulated orthorhombic  phases   ($m_1 \times m_2\times   m_3$)
when $\infty>m_1,m_2,m_3>1$.  It always higher   than that of the  phases ($1 \times
m_2\times m_3$) or ($1 \times 1\times m_3$).

\section{Mixed Lamellar phases}\label{sec:mix}
The inverse  Laplacian  approximation allows to calculate  exactly the
energy of    a    large number  of   periodic    structures   at  zero
temperature. As an illustration we present  here the results for mixed
lamellar  phases  that  are potential  ground-state  candidates in the
region of $Q/J$ where the most stable among the simple lamellar phases
involve  lamellae of width  $m=1$  (phase  $<1>$) and  $  m=2$  (phase
$<2>$).
\subsection{The $<1^n2^p>$ phases.}
The  $<1^n2^p>$  phases are the simplest  mixed   phases that  one can
construct with  the two elementary bricks formed  by lamellae of width
$m=1$ and    $m=2$:   they are formed  by     a periodic  sequence  of
ferromagnetically aligned layers  whose fundamental period consists of
$n$ one-layer   lamellae  followed  by $p$ two-layer    lamellae,  two
successive lamellae being formed by spins of  opposite signs.  Because
of the electoneutrality   (zero magnetization) condition  one has  to
distinguish   three different families:  ($n=2q$, $p=2r$),  ($n=2q+1$,
$p=2r$)  and ($n=2q$, $p=2r+1$), where $q$  and  $r$ are integers.  The
$<1^{2q+1}2^{2r+1}>$ phase is not allowed  at zero temperature because
they do not satisfy  the electoneutrality  condition.  Because of  the
two-dimensional  in-layer ferromagnetic  ordering, the    wave-vectors
characterizing lamellar phases have only one nonzero component.

\subsubsection{The $<1^{2q}2^{2r}>$ phases.}
The size of the one-dimensional unit cell is $L=2q+4r$ and the allowed
values  of the nonzero components  of the wave vector are $k=\frac{2\pi
l}{2q+4r}$ where $l$ is an integer such that $l=0,\ldots,2q+4r-1$. Summing
over all sites of the lattice, one finds
\begin{equation}
\label{eq:51}
\mid\hat{S}(k)\mid=\frac{\sqrt{N}}{2q+4r}\frac{\mid\sin(2rk)\mid}{\mid\cos(k)cos(\frac{k}{2})\mid}=
\frac{\sqrt{N}}{2q+4r}\frac{\mid\sin(qk)\mid}{\mid\cos(k)cos(\frac{k}{2})\mid},
\end{equation}
if  $k\neq  \pm  \frac{\pi}{2}$  and    $k\neq  \pi$. Otherwise,  one  obtains
$\mid\hat{S}(\pm      \frac{\pi}{2})\mid=\frac{\sqrt{N}}{2q+4r}\sqrt{8}r$ and
$\mid\hat{S}(\pi)\mid=\frac{\sqrt{N}}{2q+4r}2q$.

Using the identities
\begin{equation}\label{eq:52}
\sum_{l=1}^{q+2r-1}\frac{\sin(\frac{q\pi l}{q+2r})^2}
{\cos(\frac{\pi l}{q+2r})^2\cos(\frac{\pi l}{2(q+2r)})^2}=8r(r+q)
\end{equation}
and 
\begin{equation}\label{eq:53}
\sum_{l=1}^{q+2r-1}\frac{\sin(\frac{q\pi l}{q+2r})^2}
{\sin(\frac{2\pi l}{q+2r})^2}=r(r+q),
\end{equation}
one can express the Coulombic energy per spin as
\begin{equation}
\label{eq:54}
E_c=Q\left[\pi\frac{2q^2+16r^2+16qr}{(2q+4r)^2}-2\pi v(0)\right].
\end{equation}
The short-range contribution can be  easily calculated and the total
energy per spin is equal to 
\begin{equation}
\label{eq:55}
E_{<1^{2q}2^{2r}>}=J\left(-2+\frac{q}{q+2r}\right)+
Q\left(\pi\frac{2q^2+16r^2+16qr}{(2q+4r)^2}-2\pi v(0)\right).
\end{equation}
It is now easy to show that the above energy, whatever the strictly
positive values of  $q$ and $r$ and whatever the value of $Q/J$, cannot be
less than either the energy of the $<1>$ phase or that of the $<2>$
phase. Indeed, for the conditions
$E_{<1^{2q}2^{2r}>}\leq E_{<1>}$ 
and 
$ E_{<1^{2q}2^{2r}>}\leq E_{<2>}$
to be simultaneously satisfied, one must have  
\begin{equation}
\label{eq:56}
\frac{2q+4r}{\pi q}\leq Q/J\leq\frac{q+2r}{\pi(q+r)},
\end{equation}
which is impossible.
\subsubsection{The $<1^{n}2^{2m}>$ and $<1^{2n}2^{m}>$ phases.}
To  satisfy the requirement of the  global electroneutrality, the unit
cell  of such phases is  built  as follows.   For the  $<1^{n}2^{2m}>$
phases,    the      unit    cell      is    $\underbrace{\uparrow\downarrow     \uparrow\downarrow
\ldots\uparrow}_{n}\underbrace{\downarrow\downarrow\uparrow\uparrow \ldots\downarrow\downarrow\uparrow\uparrow}_{m}\underbrace{\downarrow  \uparrow\downarrow \ldots\downarrow
}_{n}\underbrace{\uparrow\uparrow \downarrow\downarrow \ldots\uparrow\uparrow\downarrow\downarrow}_{m}$ and the nonzero components
of   the  wave-vector are   given  by $k=\frac{2\pi  l}{(8m+2n)}$ where
$l=1\ldots8m+2n-1$. After some algebra, the Fourier  transform of the spin
variable $\hat{S}(k)$ is obtained as
\begin{equation}
\label{eq:57}
\mid\hat{S}(k)\mid=\frac{\sqrt{N}}{2n+8m}\frac{2\mid\cos(\frac{nk}{2})\mid}{\mid\cos(k)cos(\frac{k}{2})\mid}
\end{equation}
if $k\neq \pi$ and  $\mid\hat{S}(\pi)\mid=\frac{\sqrt{N}}{2n+8m}2n$.
The total energy per spin is then 
\begin{equation}
\label{eq:58}
E_{<1^{n}2^{2m}>}=J\left(-2+\frac{2n}{2n+8m}\right)+
Q\left(\pi\frac{2n^2+64m^2+32mn}{(2n+8m)^2}-2\pi v(0)\right).
\end{equation}
The $<1^{2n}2^{m}>$ phase is characterized by the sequence  $\underbrace{\uparrow\downarrow
\uparrow\downarrow \ldots\uparrow\downarrow}_{n} $$\underbrace{\uparrow\uparrow\downarrow\downarrow \ldots\downarrow\downarrow\uparrow\uparrow}_{m}\underbrace{  \downarrow\uparrow
\ldots   \downarrow\uparrow }_{n} \underbrace{\downarrow\downarrow    \ldots\uparrow\uparrow\uparrow\downarrow\downarrow}_{m}$ and  the   nonzero
components of the wave-vector are given by $k=\frac{2\pi l}{(4m+4n)}$ with
$l=1\ldots4m+4n-1$. This leads to 
\begin{equation}
\label{eq:59}
\mid\hat{S}(k)\mid=\frac{\sqrt{N}}{4n+4m}\frac{2\mid\sin(nk)\mid}{\mid\cos(k)cos(\frac{k}{2})\mid}
\end{equation}
if                                $k\neq\pm\frac{\pi}{2}$,              and
$\mid\hat{S}(\pm\frac{\pi}{2})\mid=\frac{\sqrt{N}}{4n+4m}2\sqrt{2}n$.     The
total energy per spin is finally given by 
\begin{equation}
\label{eq:60}
E_{<1^{n}2^{2m}>}=J\left(-2+\frac{n}{n+m}\right)+
Q\left(\pi\frac{32n^2+64m^2+32mn}{(4n+4m)^2}-2\pi v(0)\right).
\end{equation}
As before, there is no range of the frustration parameter $Q/J$ for which
these phases become ground states of the system.
 
\subsection{Energy of  defects in the $<2>$ phase.}\label{sec:energy-defect-2}
The issue of  the stability of the mixed  versus simple lamellar
phases can be addressed in a
different    way. Starting from the    $<2>$ phase, one can calculate the
energy change brought by inserting one or  several defects of type $<1>$
within the  periodic structure.  The creation of  one such defect
results from  the  flip of a pair of  up-down spins. This corresponds
to the simplest  excitation that one expects in  the $<2>$ phase:
$\ldots \uparrow\uparrow\downarrow\downarrow\uparrow\uparrow\downarrow\downarrow\ldots \Longrightarrow\ldots\uparrow\uparrow\downarrow\uparrow\downarrow\uparrow\downarrow\downarrow\ldots$  The resulting defective structure can  also
be viewed  as  a  $<1^{4}2^{2m}>$  phase where   $m\to \infty$. The
Coulombic energy of a $<1^{4}2^{2m}>$ structure for a cell of size
$4+4m$ is derived from Eq.~(\ref{eq:54}) and reads
\begin{equation}
\label{eq:61}
e_{c<1^{4}2^{2m}>}=(4+4m)E_{c<1^{4}2^{2m}>}=Q\left[\pi\frac{2+4m^2+8m}{1+m}-2\pi v(0)(4+4m)\right].
\end{equation}
If one subtracts the Coulombic energy of the $<2>$  phase for the same
unit  cell ($l=4+4m$)  from    the above equation,  one   obtains  the
Coulombic energy for one excitation in a $<2>$ phase. Strikingly, this
energy goes to  zero when $m$  goes to infinity,  which means that the
presence  of one  defect  in the  $<2>$ phase does   not  change at all  the
Coulombic energy of   this   phase. (This  is  not  a   result  of  the
macroscopic limit.)  Conversely, this defect  has a short-range energy
cost  that is   easily  obtained as  $E_{SR}=4$.  In  all the  region  of
frustration parameters $Q/J$ where the $<2>$ phase is more stable than
any other  simple lamellar phase, the presence  of one  defect is then
energetically unfavorable for the system.

We have also calculated the energy of  a $<2>$ phase where two defects
have been  introduced.  We  have first calculated   the energy  of the
$<2^{2m}1^{2p}2^{2n}1^{2q}>$ and $<2^{2m+2n}1^{2p+2q}>$ phases   which
are identical.  Setting $p=q=2$ whereas $n,m \to \infty  $, one obtains the
energy of the $<2>$phase in the presence of two  defects. We have also
found that  the introduction   of  the two  defects   is energetically
unfavorable  for the    system  for  any  value   of  the frustration.
Although we have not obtained a general proof, the phase-locking into
simple lamellar phases at zero temperature seems to
be well established for the model.

\begin{table}    
\caption{Coefficient of the leading term of the Coulombic
energy per site for periodic tubular structures $p=1,2,3,5,\infty $}
\begin{tabular}{ccc}

$p$ & $C(p)$ & $C(p)*p^2$\\
\tableline
 1	&	0.22208 	&	0.22208	\\
 2	&	0.089083 &	0.35633 \\
 3	&	0.045957 &	0.41361\\
 4	&	0.027568 &	0.44109\\
 5	&	0.018304 &	0.4576\\
$\infty$	&	0		& $\pi/6=0.52359\ldots$ \\
\end{tabular}

\label{tab:1}
\end{table}  

\end{document}